# Multiview Video Compression Using Advanced HEVC Screen Content Coding

Jarosław Samelak, Marek Domański

*Poznań University of Technology, Institute of Multimedia Telecommunications, Poland*
*jsamelak@multimedia.edu.pl, marek.domanski@ put.poznan.pl*

**Abstract**

The paper presents a new approach to multiview video coding using Screen Content Coding. It is assumed that for a time instant the frames corresponding to all views are packed into a single frame, i.e. the frame-compatible approach to multiview coding is applied. For such coding scenario, the paper demonstrates that Screen Content Coding can be efficiently used for multiview video coding. Two approaches are considered: the first using standard HEVC Screen Content Coding, and the second using Advanced Screen Content Coding. The latter is the original proposal of the authors that exploits quarter-pel motion vectors and other nonstandard extensions of HEVC Screen Content Coding. The experimental results demonstrate that multiview video coding even using standard HEVC Screen Content Coding is much more efficient than simulcast HEVC coding. The proposed Advanced Screen Content Coding provides virtually the same coding efficiency as MV-HEVC, which is the state-of-the-art multiview video compression technique. The authors suggest that Advanced Screen Content Coding can be efficiently used within the new Versatile Video Coding (VVC) technology. Nevertheless a reference multiview extension of VVC does not exist yet, therefore, for VVC-based coding, the experimental comparisons are left for future work.

## 1. Introduction

Virtual navigation, free viewpoint television, virtual reality or even video-based surveillance more and more rely on multiple synchronized cameras that shoot the same objects from different directions [1]. Therefore, also the compression of multiview video is gaining importance [2]. All the most important international standards on video compression, like MPEG-2 [3], AVC [4][5], and HEVC [6][7][8][9] provide multiview profiles that exploit the redundancy implied by the similarities between the views. Unfortunately, such multiview video coding requires codecs that are different from those widely used for monoscopic (single-view) video [10], therefore their practical applications are limited. An alternative solution for compression of multiview video is "frame-compatible" coding, where several views are packed into a single frame. In that way, multiview video may be compressed by a standard monoscopic encoders, similarly as for transmission of stereoscopic video for commercial 3D television [11]. The advantage is also that no side synchronization information is needed. Moreover, modern video encoders are able to compress video at very high resolutions, such as "8K" (7860×4320), which allows to accommodate multiple views (e.g. 16 views in the HD 1920×1080 format) in a single frame [6]. The major drawback of compression of frame-compatible video with a standard monoscopic encoder is that such encoder does not take advantage of the similarities between the views that compose a frame. The efficiency of such solution is roughly the same as if all the views were encoded separately [12].

In this paper, the authors present a solution for multiview video compression that has all the advantages of frame-compatible coding, but provides compression efficiency comparable to the state-of-the-art dedicated multiview video encoders.

## 2. Application of Screen Content Coding to multiview video compression

In one of previous papers [12], the authors have proposed a solution for improving the HEVC compression efficiency of the frame-compatible stereoscopic video by the use of HEVC Screen Content Coding (SCC) [13]. In the proposal, the Intra Block Copy [14] and other techniques introduced by the Screen Content Coding extension were applied to frame-compatible stereoscopic video compression to take advantage of similarities between views that compose a stereopair. Experimental results showed that application of standard Screen Content Coding in compression of frame-compatible, camera-captured stereoscopic video provides on average 15-20% bitrate reduction as compared to the simulcast compression. In another work [15], the authors investigated application of this idea to multiview video compression. For compression of 4 views, the proposal provided 20% bitrate reduction as compared to the simulcast compression. It is a significant gain, however still smaller than using dedicated multiview extension [15], due to a number of limitations and simplifications introduced by Screen Content Coding specification [6][13].

In this paper, the authors applied Screen Content Coding to multiview frame-compatible video containing natural camera-captured content, and proposed a set of improvements in order to better adapt the SCC extension to the new application. The aim of the work was to achieve compression efficiency comparable to the state-of-the-art MV-HEVC. In the context of the forthcoming Versatile Video Coding standard [16], such achievement may indicate that the Multiview and Screen Content Coding extensions could be merged into one, universal solution. For the purpose of this paper, the authors called such solution Advanced Screen Content Coding (ASCC).

## 3. Configuration of Standard Screen Content Coding

In this section, configuration of a standard SCC encoder is considered for applications in compression of frame-compatible multiview video. The goal is to exploit standard coding tools of SCC in a way that will provide better compression efficiency as compared with simulcast coding of views. For the bitstreams produced by encoders described here, full compliance with HEVC SCC specification is ensured [6].

### 3.1. Frame-compatibility

The main idea of this work is to compress multiview video as a single-layer, frame-compatible video. The creation of frame-compatible sequence was done as a pre-processing phase. The views from multiview video were joined side by side, in order middle-leftmost-rightmost (Fig. 1), which follows the coding order from Multiview HEVC Common Test Conditions [16]. The information about the order of views is proposed to be added to the bitstream as an extension of Video Parameter Set (VPS) [6]. Such information could be used after decompression of frame-compatible video to separate the views and display them in the correct order.

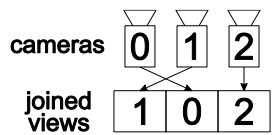

Fig. 1. Order of positioning cameras in scene and joining views – an example for 3-view video.

### 3.2. Tile-encoding

In HEVC, slices are divided into CTUs (Coding Tree Units) and, by default, encoded in rows from top left to bottom right CTU. In case of multiview frame-compatible video, it results in compressing first rows of each view, then the second rows etc. When applying Intra Block Copy, the area that can be used for matching the most similar blocks of points is limited to the previously compressed part of the slice (Fig. 2). This means that the IBC cannot use as a reference blocks of points from different view located lower than the currently analysed block. Obviously, it can reduce the efficiency of compression.

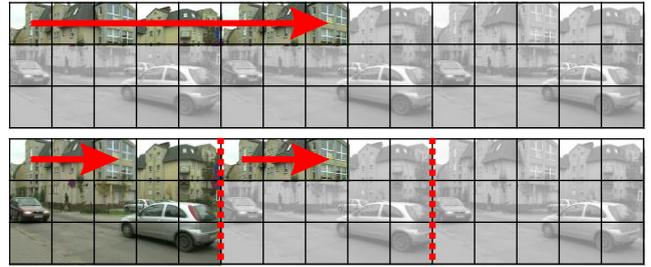

Fig. 2. Coding order without (above) and with (below) tile encoding.

In the ASCC, the encoder is configured for compression in tiles [7]. The slice is divided into three tiles, each of which contains the content from a single camera (one view). Now, the leftmost tile is entirely compressed before the encoder starts to analyse the remaining tiles. This way, whole leftmost tile (which is tantamount to the middle view) is available as a reference for Intra Block Copy applied to the remaining tiles (tantamount to the side views). Such an approach reflects the coding order in Multiview HEVC, in which the middle view is compressed as first and then it becomes a reference for compression of the remaining views.

### 3.3. Configuration of other Screen Content Coding tools

The Screen Content Coding extension is a set of tools developed for efficient compression of computer-generated content. Since the authors apply it to video containing camera-captured content, some of the tools may be inefficient. The evaluation of the SCC tools in camera-captured video compression was already presented in [12]. As a result of this work, the authors made following changes in the Screen Content Coding configuration:

- enabled Intra Boundary Filter [13],
- disabled Hash-Based Motion Estimation [13],
- disabled Palette Mode [13][18],
- disabled Colour Transform [13].

In the context of camera-captured video compression, the abovementioned modifications improve the encoding time against original SCC-HEVC with negligible influence on the efficiency.

### 3.4. Efficiency of multiview video coding using standard SCC

Section 6 reports the experimental results for compression of multiview video using standard Screen Content Coding. The results (Random Access, 3 views) demonstrate that SCC is configurable in such a way that the total bitrate may be reduced by roughly 22% as compared to simulcast HEVC with the same quality of decoded video. Nevertheless the total bitrate is higher by 25% as compared to standard MV-HEVC. In order to achieve the coding performance of MV-HEVC, some extensions of SCC have to be introduced as described in the following section.

## 4. Advanced Screen Content Coding

In the previous section, the configuration of standard Screen Content Coding was considered for compression of frame-compatible multiview video. In this section, the authors propose a set of modifications for improving the compression efficiency of multiview video. These improvements are not compatible with the original SCC, but they aim at achieving compression efficiency comparable to the Multiview HEVC, which is the state-of-the-art dedicated encoder for compression of multiview video.

### 4.1. Intra Block Copy vectors precision

The output of the Intra Block Copy tool is a vector that points from the currently analysed block of points to the most similar block within the same picture. In the state-of-the-art IBC, such vector has a full-pel precision. The authors introduced to IBC a quarter-pel precision to follow the precision of disparity vectors in Multiview HEVC. Such a solution improves the accuracy of the prediction, especially for the camera-captured multiview video. On the other hand, the quarter-pel accuracy is redundant in compression of single view video that presents screen content. The influence of the introduced changes on the compression efficiency of screen content was investigated in Section 7.

### 4.2. Starting point for block matching in Intra Block Copy

The goal of using Intra Block Copy, frame-compatibility and tile encoding is to match blocks of points from one view with corresponding blocks of points from another view. The resulting vector is expected to be very long, since it will point to a different tile (Fig. 3). Obviously, compression of long vectors is less efficient than short ones. Additionally, it will take a lot of time for the Intra Block Copy to find the optimal match because it starts the search from the area nearby the analysed block [14].

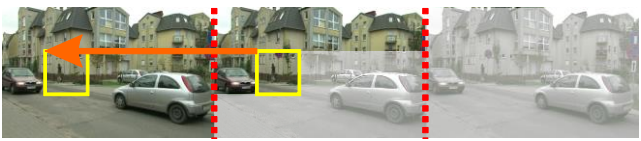

Fig. 3. Long vector derived by Intra Block Copy.

In the proposed solution, the starting point for the IBC is the position of the collocated block in the leftmost tile, which reduces the length of the resulting vector, as well as the time to find it. Additionally, the authors replaced the state-of-the-art IBC search algorithm with the one used in Motion Compensated Prediction, as it appeared to be more efficient for the camera-captured content.

### 4.3. SAO and deblocking filtering per tile

At the end of compression of each slice, the HEVC encoder applies two filters to the reconstructed image: SAO (Sample Adaptive Offset) and deblocking filter [7]. They are intended to increase the quality of the reconstructed image by reducing the influence of the encoding artefacts: ringing and block effect. In frame-compatible multiview video compression, these filters would be applied after compression of each three views, because they compose a single slice. Therefore, the leftmost tile would suffer from the encoding artefacts at the time it was used as a reference for compression of the remaining tiles, resulting in lower compression efficiency. Because of that, the proposed encoder applies SAO and deblocking filter after compression of each tile. Additionally, filtering across tile boundaries was disabled, because the sharp edge between views is expected and should be preserved.

### 4.4. Different Quantization Parameter for side views

In Multiview HEVC, the quality of the reconstructed video can be controlled separately for each view using Quantization Parameters (QPs). In the frame-compatible solution, all the views compose a single slice and therefore they cannot be encoded at different QPs. The proposed improved SCC-HEVC codec was equipped with the possibility to define different Quantization Parameter for each tile. The information about applying different QP to the side tiles was included in the bitstream within the VPS extension. It is encoded as difference between original QP (applied to the leftmost tile), and the desired QP for the remaining tiles.

### 4.5. Reference tile border extension

After compression of the first view, Multiview HEVC encoder saves the reconstructed image to use it later as a reference in compression of the remaining views. The borders of such image are extended to enable the Motion Compensated Prediction to effectively search for the most similar block of points close to the borders of the reference image [7].

In the proposed solution, prediction of blocks close to the right border of a tile may be inaccurate because of the border between the reference (leftmost) tile and its neighbour (Fig. 4). In order to avoid this issue, the reconstruction of a reference tile is separated from the whole image and its borders are extended. Such a solution improves the accuracy of the prediction at the borders between tiles.

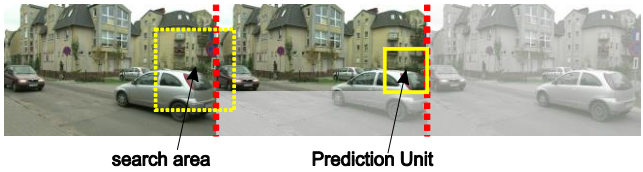

Fig. 4. Problem at the right borders of tile.

## 5. Description of the experiments

The goal of modifications described in Section 4 was to achieve compression efficiency similar to MV-HEVC, which is the dedicated codec for compression of multiview video. Therefore, a number of experiments were conducted to compare the efficiency of MV-HEVC against original and ASCC. The encoders were compiled from the corresponding reference software, according to Table 1. The ASCC improvements were implemented on top of the HEVC-SCC codec.

Table 1. Used encoders and corresponding software.

| Encoder | Software |
|---|---|
| MV-HEVC | HTM-16.2 [19] |
| HEVC (simulcast) | HM-16.9 [20] |
| HEVC-SCC (frame-compatible) | HM-16.9+SCM-8.0 [21] |
| ASCC (frame-compatible) | HM-16.9+SCM-8.0 + authors' improvements |

All codecs are based on the same version of HEVC (HM-16.9 [20]), therefore the results are not influenced by any differences other than implemented improvements, Screen Content Coding or Multiview extension. The encoders were configured with respect to the appropriate Common Test Conditions [22][23][17], with some changes in the MV-HEVC configuration, in which the vertical disparity search range was set to 64 and Early Skip Detection was turned on to make it consistent with HEVC-SCC configuration. Obviously, the configuration of ASCC was also modified according to the improvements proposed in Section 3.

The tests were performed on 2 and 3 views obtained from 8 commonly used multiview sequences [24][25][26][27]. The chosen sequences, views and order of views were compliant with the Common Test Conditions for compression of multiview video [17]. The views in these sequences are in parallel and contain both camera-captured and computer-generated content. The experiments were conducted in two different coding scenarios: All Intra (only intra-frame and inter-view prediction allowed) and Random Access (inter-frame prediction allowed as well, intra period equal to 24). Each time, both compression efficiency and encoding time were measured. All experiments were performed on a PC with Intel Xeon 3GHz CPU.

## 6. Experimental results for multiview video

This section presents the outcome of comparison of original Screen Content Coding and Advanced SCC against MV-HEVC and simulcast HEVC. Tables 2 and 3 contain experimental results of encoding in All Intra and Random Access coding scenarios, respectively. The compression efficiency is presented as an average bitrate alteration, calculated according to the Bjøntegaard metric for luma PSNR [28]. Values below zero indicate that the given encoder provides lower bitrate (or shorter encoding time) than the reference MV-HEVC encoder (or simulcast HEVC).

Table 2. Experimental results against HEVC simulcast and MV-HEVC – All Intra

| Sequence | HEVC simulcast | | | | | | | | MV-HEVC | | | | | | | |
|---|---|---|---|---|---|---|---|---|---|---|---|---|---|---|---|---|
| | 2 views | | | | 3 views | | | | 2 views | | | | 3 views | | | |
| | Bitrate [%] | | Enc time [%] | | Bitrate [%] | | Enc time [%] | | Bitrate [%] | | Enc time [%] | | Bitrate [%] | | Enc time [%] | |
| | SCC | ASCC | SCC | ASCC | SCC | ASCC | SCC | ASCC | SCC | ASCC | SCC | ASCC | SCC | ASCC | SCC | ASCC |
| Poznan Hall 2 | -16.79 | -28.01 | +56 | +8 | -21.08 | -40.57 | +63 | +22 | 17.24 | 2.93 | +43 | -1 | 27.36 | -1.67 | +32 | -5 |
| Poznan Street | -20.66 | -31.57 | +170 | +38 | -27.53 | -47.35 | +183 | +48 | 15.30 | 2.69 | +79 | -9 | 29.05 | -0.55 | +67 | -12 |
| Kendo | -20.25 | -29.19 | +169 | +117 | -26.33 | -40.76 | +183 | +147 | 14.10 | 2.19 | +25 | +1 | 22.34 | 0.06 | +11 | -4 |
| Balloons | -21.49 | -29.29 | +195 | +121 | -30.08 | -41.90 | +212 | +148 | 12.01 | 1.80 | +35 | +1 | 18.27 | 0.00 | +24 | -1 |
| Newspaper | -18.32 | -25.37 | +242 | +131 | -25.51 | -39.06 | +259 | +151 | 9.75 | 1.87 | +51 | +2 | 18.56 | -0.35 | +37 | -3 |
| Undo Dancer | -35.14 | -39.98 | +330 | +30 | -47.16 | -55.71 | +319 | +31 | 8.86 | 1.25 | +224 | -2 | 17.12 | -0.79 | +204 | -5 |
| GT Fly | -38.56 | -42.27 | +150 | +11 | -50.71 | -58.22 | +126 | +4 | 6.85 | 0.62 | +124 | +0 | 15.36 | -1.54 | +111 | -3 |
| Shark | -35.39 | -40.50 | +180 | +35 | -49.78 | -57.29 | +142 | +27 | 8.89 | 0.64 | +111 | +2 | 15.68 | -0.97 | +87 | -2 |
| Average | -26.28 | -34.21 | +175 | +41 | -34.56 | -48.52 | +175 | +50 | 12.47 | 1.94 | +99 | -2 | 22.25 | -0.90 | +85 | -5 |

Table 3. Experimental results against HEVC simulcast and MV-HEVC – Random Access

| Sequence | HEVC simulcast | | | | | | | | MV-HEVC | | | | | | | |
|---|---|---|---|---|---|---|---|---|---|---|---|---|---|---|---|---|
| | 2 views | | | | 3 views | | | | 2 views | | | | 3 views | | | |
| | Bitrate [%] | | Enc time [%] | | Bitrate [%] | | Enc time [%] | | Bitrate [%] | | Enc time [%] | | Bitrate [%] | | Enc time [%] | |
| | SCC | ASCC | SCC | ASCC | SCC | ASCC | SCC | ASCC | SCC | ASCC | SCC | ASCC | SCC | ASCC | SCC | ASCC |
| Poznan Hall 2 | -11.30 | -23.55 | +80 | -5 | -13.82 | -32.09 | +77 | -2 | 13.47 | -1.92 | +88 | -1 | 25.43 | -0.63 | +80 | +0 |
| Poznan Street | -13.88 | -25.78 | +141 | +22 | -19.29 | -38.91 | +140 | +23 | 16.73 | 1.43 | +102 | +2 | 30.82 | 0.51 | +99 | +2 |
| Kendo | -10.30 | -19.45 | +169 | +45 | -13.80 | -27.52 | +127 | +47 | 12.37 | 1.45 | +88 | +1 | 18.39 | 0.37 | +56 | +1 |
| Balloons | -13.26 | -21.38 | +190 | +45 | -18.44 | -30.11 | +183 | +45 | 11.28 | 1.34 | +100 | +0 | 16.19 | 0.33 | +92 | -1 |
| Newspaper | -16.00 | -23.55 | +184 | +47 | -20.46 | -33.04 | +189 | +50 | 8.93 | -0.03 | +97 | +2 | 17.74 | 0.46 | +94 | +0 |
| Undo Dancer | -25.82 | -35.02 | +175 | +17 | -34.32 | -47.59 | +176 | +17 | 11.89 | -1.69 | +137 | +1 | 24.67 | 0.05 | +138 | +1 |
| GT Fly | -21.74 | -31.09 | +162 | +10 | -31.67 | -46.23 | +142 | +7 | 13.87 | 0.21 | +134 | -2 | 27.18 | 0.05 | +123 | -1 |
| Shark | -29.50 | -37.76 | +164 | +30 | -40.39 | -52.22 | +123 | +28 | 12.82 | -0.10 | +104 | +0 | 23.93 | -0.18 | +74 | +0 |
| Average | -16.61 | -26.98 | +145 | +18 | -22.58 | -38.47 | +132 | +18 | 13.67 | -0.11 | +110 | +0 | 25.30 | 0.07 | +99 | +0 |

The results prove that the ASCC is much more efficient than original HEVC-SCC for frame-compatible multiview video compression, and roughly as efficient as MV-HEVC. In the All Intra encoding scenario, ASCC is the most efficient solution. First of all, it includes in the bitstream less slice headers, compared to the MV-HEVC. Secondly, in the proposed solution some signalization data related to the multiview layer may be omitted, e.g. the number of views is equal to the number of tiles, thus there is no point to duplicate this information. In the Random Access encoding scenario, ASCC is slightly less efficient, mostly due to differences between HEVC-SCC and MV-HEVC in creating reference pictures lists. In case of HEVC-SCC, the current picture is simply the last reference picture on the first reference pictures list, while in MV-HEVC, the side views are put on the reference pictures lists in a more sophisticated manner.

## 7. Experimental results for screen content video

The previous section presents the comparison between MV-HEVC and ASCC for compression of multiview video. In this section, the authors assess the influence of the introduced modifications on the compression efficiency of the single view screen content video. Tables 4 and 5 present the results for Advanced SCC against unmodified Screen Content Coding. The configuration of both encoders was exactly the same and compliant with Common Test Conditions [23]. The sequences used in the experiment were also recommended in CTC and they contain computer-generated content. As previously, the experiment was conducted in two coding scenarios: All Intra and Random Access.

Table 4. Experimental results against HEVC-SCC – 1 view screen content.

| Sequence | All Intra | | Random Access | |
|---|---|---|---|---|
| | Bitrate [%] | Enc time [%] | Bitrate [%] | Enc time [%] |
| Basketball_Screen | 3.90 | +22 | 2.60 | +10 |
| ChinaSpeed | 0.58 | +14 | 0.32 | +23 |
| ChineseEditing | 3.95 | +13 | 3.50 | +12 |
| MissionControlClip2 | 0.56 | +19 | 0.42 | +10 |
| MissionControlClip3 | 3.05 | +20 | 2.38 | +10 |
| sc_console | 15.07 | +30 | 8.34 | +15 |
| sc_desktop | 13.40 | +23 | 8.82 | +13 |
| sc_flyingGraphics | 9.07 | +20 | 5.38 | +31 |
| sc_map | 2.35 | +16 | 1.42 | +10 |
| sc_programming | 4.61 | +18 | 1.89 | +13 |
| sc_robot | 0.28 | +13 | 0.10 | +20 |
| sc_web_browsing | 15.24 | +26 | 11.27 | +18 |
| SlideShow | -0.98 | +23 | -0.77 | +33 |

Considering compression of only one view as a single tile, most of the authors' modifications described in sections 3 and 4 did not affect the compression process. The only change that influenced the results was the quarter-pel accuracy of the Intra Block Copy vectors. Such modification obviously increased the encoding time, but the compression efficiency usually did not benefit from it. The authors observed that the results vary a lot and they strongly depend on the content of the sequence. The compression efficiency of ASCC is close to the unmodified SCC for sequences that contain a lot of fluent motion, gradients, parts with camera-captured content, or computer-generated images that are supposed to imitate natural images. On the other hand, the bitstream can increase even by 15% if the sequence is mostly composed of text, simple graphics and simple motion. In such cases, the encoder does not benefit from quarter-pel accuracy of Intra Block Copy vectors, but more bits are used to transmit them. In order to avoid this redundancy, the precision of the IBC vectors could be dynamically adjusted by the encoder, depending on the content of the video

## 8. Conclusions

In the paper, the authors proposed an efficient frame-compatible multiview video compression technique that utilizes HEVC Screen Content Coding extension. A number of modifications of the state-of-the-art HEVC-SCC were introduced, which resulted in achieving solution roughly as efficient as the dedicated multiview extension. For compression of screen content, the proposed modifications usually cause a surplus in the bitstream, but it strongly depends on the content of the video and could be reduced by choosing the Intra Block Copy vectors precision adaptively. The advantage of authors' proposal is that Screen Content Coding has more applications and does not require multi-layer encoder and special infrastructure. Therefore, with little effort, the SCC extension can be successfully reused for compression of multiview video. In the context of the forthcoming Versatile Video Coding standard [16], the authors claim that the Multiview and Screen Content Coding extensions should not be developed separately, but as a single, universal solution.